\documentclass[aps,pre,twocolumn,groupedaddress]{revtex4}
\usepackage{graphicx,amsmath}
\begin{document}

\title{Why is Understanding Glassy Polymer Mechanics So Difficult?}
\author{Robert S. Hoy}
\affiliation{Departments of Mechanical Engineering \& Materials Science, and Physics, Yale University, New Haven, CT} 
\date{\today}
\begin{abstract}
In this Perspective, I describe recent work on systems in which the traditional distinctions between \textbf{(i)} unentangled vs.\ well-entangled systems and \textbf{(ii)} melts vs.\ glasses seem least useful, and argue for the broader use in glassy polymer mechanics of two more dichotomies: systems which possess \textbf{(iii)} unary vs.\ binary and \textbf{(iv)} cooperative vs.\ noncooperative relaxation dynamics.  
I discuss the applicability of \textbf{(iii-iv)} to understanding the functional form of strain hardening.
Results from molecular dynamics simulations show that the ``dramatic'' hardening observed in densely entangled systems is associated with a crossover from unary, noncooperative to binary, cooperative relaxation as strain increases; chains stretch between entanglement points, altering the character of local plasticity.
Promising approaches for future research along these lines are discussed.
\end{abstract}
\maketitle

\begin{figure}[htbp]
\includegraphics[width=3.1in]{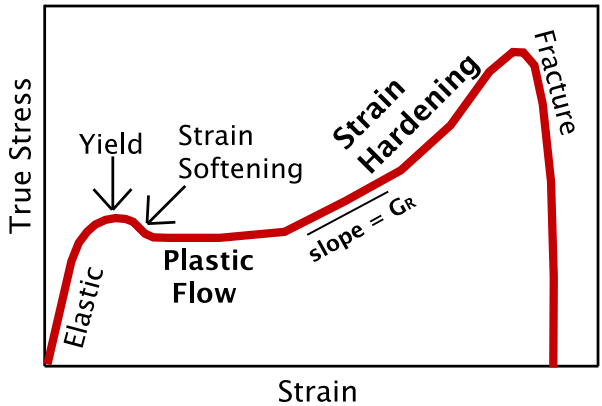}
\caption{Schematic of stress-strain curves for ductile polymer glasses.  Dramatic hardening coincides with the increase in slope at large strains.  In brittle systems, fracture intervenes at lower strains because strain hardening is insufficient to stabilize the material against post-yield strain localization \cite{haward97}.}
\label{fig:GSHschematic}
\end{figure}

One of the reasons dense polymeric systems so interest physical scientists,  apart from their ubiquity and utility, is the wide range of energy, length, and time (EL\&T) scales controlling their properties.
Figure \ref{fig:GSHschematic}, a schematic depiction of typical stress-strain curves for ductile polymer glasses, illustrates the mechanical consequences of this range.  
Undeformed systems occupy low-lying regions of a rugged free energy landscape.
In the linear elastic regime, systems remain near initial free energy minima, and stress is controlled by local forces at the Kuhn scale or below.
Yield occurs when energetic barriers to segmental rearrangements are overcome; the resulting increase in local mobility produces strain softening.
In the plastic flow regime, the stress $\sigma = \partial W/\partial \epsilon$ is relatively constant.
Strain hardening begins when $\partial W/\partial \epsilon$ must increase to drive further segmental rearrangements while maintaining chain connectivity.
This increase becomes more dramatic as the scale over which chains are oriented approaches that of the entanglement mesh.
Finally, fracture occurs when cohesive forces, either primary covalent bonds or secondary nonbonded interactions, no longer suffice to maintain material integrity.

The industrial importance of understanding strength and failure of polymeric materials has made quantitatively predicting the entire range of mechanical response shown in Fig.\ \ref{fig:GSHschematic} one of the main goals of physical polymer science. 
While great progress has been made in recent years towards understanding phenomena at strains up to and including the early stages of strain hardening within a single framework, a coherent theoretical picture including dramatic hardening and fracture remains elusive. 
In this Perspective I discuss some of the reasons why this is so, and outline possible strategies for moving forward.

Polymer dynamics and mechanics are often cast in terms of polar dichotomies; a given system is classified as
belonging to one ``pole'' or the other.
The two most commonly employed dichotomies for bulk amorphous systems (at least those relevant to Fig.\ \ref{fig:GSHschematic}) are those between: \textbf{(i)} unentangled vs.\ well-entangled systems, and \textbf{(ii)} melts far above $T_g$ vs.\ glasses far below $T_g$.
Dynamical behavior in polar limits (e.\ g.\ unentangled melts \cite{doi86}) is amenable to relatively simple theoretical treatment.
Because the EL\&T scales controlling relaxation are well separated, one may focus on a dominant scale and then treat relaxation on that scale.
In the context of Fig.\ \ref{fig:GSHschematic}, only glassy systems posess a finite plastic flow stress, and only entangled systems strain harden.
Flow and hardening have therefore traditionally been explained in terms of local plasticity \cite{boyce88} and the work required to deform the entanglement network \cite{arruda93b}, respectively.

Dichotomies \textbf{(i-ii)} have obviously proven of great utility for understanding polymer mechanics, but for actively deforming systems, they are less useful. 
A classic example is the well known tendency of systems \textit{above} their quiescent $T_g$ to exhibit glass-like mechanical features when deformed sufficiently rapidly. 
Significant strain hardening occurs when the product of strain rate $\dot\epsilon$ and chain-scale relaxation time $\tau$ exceeds unity.
More recent experimental, theoretical and simulation work \cite{loo,chen,lee} has further blurred the distinction between melts and glasses by showing that systems \textit{below} their quiescent $T_g$ become meltlike on short length scales under active deformation.
Finite strain rates can reduce the segmental relaxation time $\tau_{\alpha}$ by orders of magnitude; $\dot{\epsilon}\tau_{\alpha}$ drops well below unity at yield.
This drop is partially reversed in the strain hardening regime, and in all cases, $\tau_{\alpha}$ increases dramatically when deformation is ceased. 
Clearly the terms ``meltlike'' and ``glassy'' poorly characterize such nonlinear behavior.
Categorizing systems as either ``entangled'' or ``unentangled'' can be similarly complicated, because entanglements (unlike crosslinks) may be short-lived or long-lived compared to experimental time scales, and their relaxation is also altered by active deformation.
These developments have helped clarify why developing a coherent theoretical framework for predicting the entire range of behavior depicted in Fig.\ \ref{fig:GSHschematic} is so difficult.

How, then, to proceed?  
The great advances made by conceptualizing mechanical properties in terms of dichotomies \textbf{(i-ii)} suggest that looking for additional ones is a useful strategy. 
It seems to me that two potentially very useful dichotomies for improving our understanding of glassy polymer mechanics are: \textbf{(iii)} unary vs.\ binary relaxation and \textbf{(iv)} noncooperative vs.\ cooperative relaxation.
Like \textbf{(i)} and \textbf{(ii)}, \textbf{(iii)} and \textbf{(iv)} are dichotomistic views of the \textit{character} of the dominant relaxation mechanisms.
The ``poles'' correspond to whether the constituents of a system can be accurately \textit{treated} as relaxing independently of one another.
If relaxation is unary and/or noncooperative, they can.
If it is binary and/or cooperative, they cannot; instead, one must explicitly treat correlations between constituents.
In systems with noncooperative relaxation, a tracer particle would have the same motion in a system where all other particles are frozen as in its unfrozen counterpart \cite{foffi00}.
In systems with cooperative relaxation, the opposite is true.

An example illustrating the unary vs.\ binary dichotomy is as follows; suppose the stress $\sigma$ is a function of the intra-chain statistics $\left<R^{2}(n)\right> = |\vec{r}_{i} - \vec{r}_{i+n}|^2$, and perhaps the history of $\left<R^{2}(n)\right>$, where $n$ is chemical distance.  
 In other words, suppose the stress is controlled by chain configuration but interchain correlations are either unimportant or trivially integrable.
Then the stress relaxation processes are unary.  
The classical theory of rubber elasticity, which assumes conformations of single strands at the scale $n = N_c$ (the distance between crosslinks) control stress, is a unary-relaxation theory.
On the other hand, suppose local interchain orientation is important; relaxation processes will then be  binary or higher order.

A paradigmatic higher-order process in polymers is disentanglement, which occurs when one chain end deintersects another chain.
This is a \textit{binary} process because it involves two chains.
While it is often useful to approximate disentanglement as an ``infinite'' order process and ``wrap'' it into a unary mean-field theory (\textit{i.\ e.}, the tube theory of melt dynamics \cite{doi86}), it is well known that in many cases such a description becomes inadequate  \cite{mcleish02}.
In principle, binary processes require a formal description that utilizes some two point correlation function $F(\vec{r}_1, \vec{r}_2)$ where $\vec{r}_1$ and $\vec{r}_2$ lie on different chains.

Dichotomies \textbf{(iii-iv)} are are not new concepts, and have previously been applied in polymer melt rheology \cite{mcleish02} as well as many other fields.
They are subject to ambiguities similar to those mentioned above for  \textbf{(i-ii)}; the same system may have unary and binary and / or cooperative and noncooperative relaxation processes occurring simultaneously on different length scales.
Many processes are known to become increasingly cooperative as $T_g$ is approached from above \cite{ediger00}, and remain so below $T_g$, but predicting the \textit{degree} of cooperativity is difficult.
Another useful measure of cooperativity is the degree of coupling between \textit{different} relaxation processes. 
This clearly strengthens as separation of the relevant  EL\&T scales decreases, but quantitative prediction is challenging.
Furthermore, in systems exhibiting nonlinear response, it is typically \textit{a priori} unclear which couplings are most important in which regimes.
Such issues remain controversial even for simple (e.\ g.\ metallic and colloidal) glassformers.
However, since few theories of glassy polymer mechanics explicitly consider them, opportunities for progress abound.
I now give an example illustrating how even \textit{qualitative} treatment of dichotomies \textbf{(iii-iv)} can elucidate the nature of the crossover to dramatic hardening.

Understanding the role played by entanglements in controlling the mechanical properties of polymer glasses has proven extremely difficult.
The canonical \cite{haward93} functional form of stress-strain curves at large strains, ``Gaussian'' strain hardening, is given by 
\begin{equation}
\sigma = \sigma_0 + G_R g(\bar{\lambda}),
\label{eq:gaussian}
\end{equation}
where $\sigma_0$ is a flow stress, $G_R$ is the strain hardening modulus, $\bar\lambda$ is the macroscopic stretch tensor, and $g(\lambda)$ is alternately the (negative) derivative of the entropy density of an affinely stretched entanglement network, or a Neohookean term.
The traditional entropic treatment of strain hardening in glasses \cite{arruda93b} qualitatively captures the shape of stress-strain curves, but has several flaws, extensively discussed in the literature (e.\ g.\ \cite{vanMelick03,kramer05,hoy08}).

\begin{figure*}
\includegraphics[width=5.92in]{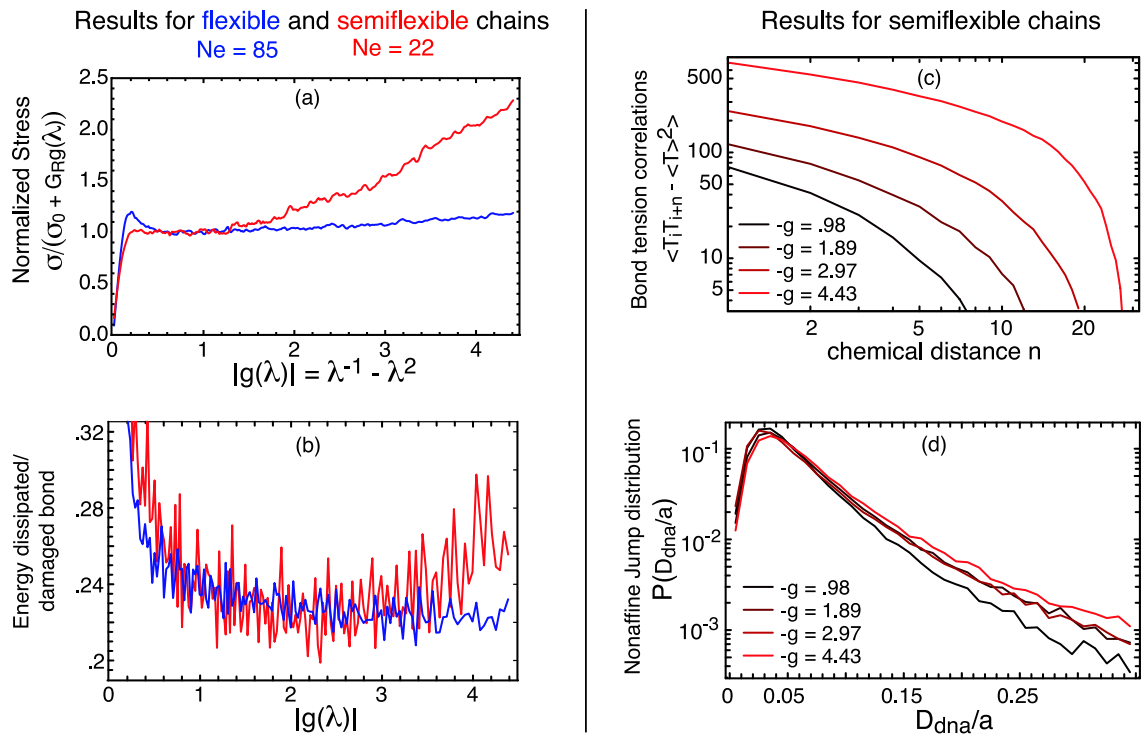}
\caption{Crossover from unary-noncooperetative to binary-cooperative relaxation in polymer glasses arising from stretching of entanglement network.  Systems are uniaxially compressed to a true strain $\epsilon = -1.5$.: $\lambda \equiv \textrm{exp}(\epsilon)$,  $|g(\lambda)| \equiv |\lambda^{-1}-\lambda^{2}|$, and $|g(-1.5)| = 4.43$. 
Simulations are performed at low temperature $T \sim T_g/35$ to minimize thermal noise, and a high strain rate $|\dot\epsilon| = 10^{-3}/\tau_{LJ}$ is purposefully chosen \cite{footm}.
All units and protocols are desribed in the Appendix.
Panels (a-b) contrast results from loosely and tightly entangled systems for (a) scaled stress $\sigma/(\sigma_0 + G_Rg(\lambda))$ and (b) energy dissipated  (in units of $u_0$) per `damaged' LJ bond.
Statistical noise in panel (b) arises from finite system-size effects.
Panels (c-d) show results for tightly entangled systems: variations with strain of (c) bond tension correlation fluctuations along chain backbones, and (d) the probability distribution for nonaffine jump sizes.}
\label{fig:btc}
\end{figure*}

Much progress has been made in the five years since Kramer challenged \cite{kramer05}  the polymer physics community to resolve this issue.
Most physically, the mechanisms underlying strain hardening in glasses have been shown to be largely viscous and closely connected to plastic flow \cite{hoy08,govaert08}.  
Yet the underlying problem of distinguishing between models which assign different roles to entanglements yet make similar predictions remains unsolved.
Theories which assume entanglements are all-important, and theories which assume they play no role, both predict the Gaussian form.
It is seemingly predicted by practically \textit{any} theory which predicts chains deform affinely on large scales, whether based on linearized entropic elasticity, Neohookean viscoplasticity \cite{hoy10} or alteration of interchain ordering and suppression of density fluctuations \cite{chen}.
Similarly, sub-Gaussian hardening, i.\ e.\ $\sigma$ sublinear in $g$, is produced by subaffine large-scale deformation, whether this arises from relaxation of the  entanglement network (as in melts or transient networks) or finite $\dot\epsilon \tau$ in uncrosslinked glasses \cite{vanMelick03,hoy10}.
Dramatic hardening  ($\sigma$ supralinear in $g$) can be produced either by entropic depletion of configurations for finite-$n$ chain segments \cite{arruda93b} or by the increased plastic deformation and bond stretching required to deform chains affinely while maintaining their connectivity.  

Formulation of a robust microscopic theory predicting the entire range of mechanical response shown in Fig.\ \ref{fig:GSHschematic} seems doubtful while these ambiguities remain.
One approach to resolving them is to connect changes in macroscopic mechanical response with increasing strain to changes in relationships between structural features and relaxation mechanisms at different length scales. 
Figure \ref{fig:btc} presents results for the crossover to dramatic strain hardening, which I will argue can best be understood as representing crossovers from unary to binary and from noncooperative to cooperative relaxation phenomena.
Results are obtained via molecular dynamics simulations of a simple coarse-grained bead-spring model \cite{kremer90} that captures the key physics of linear homopolymers.
The simulation protocol is standard \cite{hoy08} and is described in the Appendix.
Both flexible and semiflexible chains are studied to illustrate the behavior of ``loosely'' and ``tightly'' \cite{everaers04} entangled systems, respectively.

Panel (a) compares the normalized stresses ($\sigma/(\sigma_0 + G_Rg(\lambda)$)), where $G_R$ is fit to the initial hardening regime ($.5 < |g(\lambda)| < 1$). 
The plateau for flexible chains indicates nearly Gaussian hardening at all strains.
In contrast, semiflexible chains show dramatically supra-Gaussian hardening for $|g(\lambda)| \gtrsim 2$. 
Panel (b) shows the energy dissipated per damaged bond $U_d$ for the same systems.   
$U_d = \sigma^{Q}/P$, where $\sigma^{Q}$ is the dissipative component of the stress and $P$ is the rate of bond damage per unit strain (see the Appendix). 
Flexible chains again show a plateau, indicating $U_d$ is constant when hardening is Gaussian, while semiflexible chains show an increase in $U_d$ for $g(\lambda) \gtrsim 3$. 
Differences between loosely and tightly entangled systems are directly associated \cite{hoy08} with differing degrees of increase in the energetic component of stress, $\sigma^{U} = \sigma-\sigma^{Q}$.

Panel (c) shows the correlation in bond tensions 
$T = \partial U_{FENE}/\partial \ell$ along chain backbones in tightly entangled systems.
The correlations are roughly exponential. 
At intermediate $n$,  $\left<T_{i}T_{i+n}\right> \sim exp(-2n/N_e)$ in the limit of large strains; 
the factor of $2$ suggests a binary suppression of relaxation, wherein tension is concentrated at increasingly localized entanglement points (with 2 chains/entanglement) and decorrelates between entanglement points.
In contrast, bond tensions in flexible systems (not shown) are much smaller and less correlated.
This is of interest since chain tension relief by covalent bond scission is a key mechanism leading to brittle 
fracture \cite{haward97, footm}.

Panel (d) shows the probability distribution for differential nonaffine jump sizes $D^{2}_{dna}$ in tightly entangled systems, defined by
\begin{equation}
D^{2}_{dna} = \left|\vec{r}_{k+1} - \displaystyle\frac{\bar\lambda_{k+1}}{\bar\lambda_{k}}\vec{r}_{i}\right|
\label{eq:d2dna}
\end{equation}
where $\vec{r}_{k}$ is the position of a particle and $\bar\lambda_{k}$ is the macroscopic stretch tensor at $|\epsilon| = k\delta\epsilon$.
The tails of $P(D^{2}_{dna})$ become longer with increasing stress and strain \cite{warren10b}.
This effect is most pronounced for tightly entangled chains in the dramatic hardening regime (results for loosely entangled chains are similar to those found in Ref.\ \cite{warren10b} and are not shown).
It arises because correlated bond tension at the scale $n \sim N_e$ increases the size of local plastic rearrangements, which in turn increases the energy they dissipate (panel b).

All results in Fig.\ \ref{fig:btc} are consistent with crossovers from unary to binary relaxation as chains stretch between entanglements, and from noncooperativity to cooperativity between deformation at the level of the entanglement mesh and local plastic rearrangements at the monomer or Kuhn scale.
For the strains considered here, these crossovers are present in tightly entangled but not loosely entangled systems; all coincide with increasing stretching of chains over chemical distances $n\sim N_e$.
Note that the crossovers in Fig.\ \ref{fig:GSHschematic}(a-b) occur at different strains, while those in panels (c-d) are gradual.
It seems likely that the unary-binary and noncooperative-cooperative crossovers are themselves coupled, though the strength of the coupling remains unclear. 
Understanding such behavior at a predictive level may be difficult, but presents an interesting challenge for the community.

Single-chain-in-mean-field descriptions of strain hardening \cite{hoy10,chen,arruda93b} probably cannot quantitatively treat the ``entanglement-stretching'' unary-binary crossover, even if interchain correlations are integrated into the mean field.
Similarly, theories assuming a single relaxation mechanism (e.\ g.\ segmental relxn.) cannot treat crossovers from noncooperative to cooperative relaxation which are driven by coupling to another relaxation mechanism of different character.
Presently, the microscopic theory which most satisfactorily captures the elastic, yield, softening, flow, and hardening regimes is due to Chen and Schweizer \cite{chen};
it is based on a strain- and thermal-history-dependent dynamical free energy for segmental rearrangements in the glassy state.  
While it captures much of the physics of strain hardening (e.\ g.\ its coupling to plastic flow) by predicting how segmental relaxation is suppressed due to changes in interchain ordering in a macroscopic strain field, and even predicts variations in the character of activated segmental hops, it is questionable whether it can capture the crossovers shown in Fig.\ \ref{fig:btc}(b-c), and it does not treat fracture.

An interesting feature of the unary vs.\ binary dichotomy is that unary relaxation processes ``scale'' linearly with the density $\rho$ of relaxing constituents, while binary processes scale quadratically with $\rho$.
For example, the reason polymeric entanglement is traditionally regarded as a (nearly) binary process \cite{graessley81} is that entanglement density scales roughly as $\rho^{2}$, where $\rho$ is the density of uncrossable chain contours.
Studies of bidisperse systems \cite{hoy09b} (with constituent densities ($\rho_1$, $\rho_2$) and relaxation processes which scale as $\rho_1 \rho_2$) should therefore be particularly useful in testing whether relaxation is unary or binary.
Many such studies have been performed for melts well above $T_g$, in which the coupling between relaxation of short and long chains is now well understood \cite{ylitalo91}, but similar studies in systems below $T_g$ are in their infancy.
New experimental techniques such as scanning near field optical microscopy \cite{ube09} should be particularly useful for these purposes.

A specific example of how such studies can elucidate causal relationships for polymer mechanics  is as follows.
Experiments performed on (essentially) monodisperse systems remain contradictory on such basic questions as to whether or not $G_R$ is proportional to entanglement density $\rho_e$ \cite{haward93,vanMelick03,govaert08}; if it is, it is clear that the constant of proportionality differs for chemically different polymers.
In monodisperse (but not bidisperse) systems, both $G_R$ and entanglement density $\rho_e$ scale approximately as $l_K^{3}$ \cite{kramer05,fetters99b}.
This ``macro-micro'' ambiguity led to much controversy over whether $G_R$ scales fundamentally with entanglement density or Kuhn length.  
For systems exhibiting \textit{Gaussian} hardening, an apparent resolution of these issues has been obtained via recent simulation and experimental studies of bidisperse systems \cite{hoy09b,hoy10} that suggest structure at the Kuhn level is the more fundamental controlling factor.
However, the ambiguity remains unresolved in the dramatic hardening regime, and the crossovers described above suggest that developing a theoretical description that captures the entire spectrum of behavior depicted in Fig.\ \ref{fig:GSHschematic} will require a great deal more work.

In this effort, many of the most fruitful ideas may come from glass \textit{transition} physics \cite{ediger00}, which I believe has been underutilized to date in theories of glassy polymer mechanics (at least those which treat large strains), except that of Ref.\ \cite{chen}.
It may also be profitable to employ concepts developed in recent studies of other ``soft'' systems, such as colloidal and granular materials.
One promising possibility is to determine whether and when the properties of polymers are more like attractive glasses or repulsive glasses \cite{foffi00,footattrepg}. 
Other potentially useful concepts include jamming, soft modes, inherent structures and energy landscapes \cite{ohern03}.
Much also may be gained by working to bridge microscopic theories with increasingly sophisticated new constitutive models; some especially promising models incorporate melt-like relaxation mechanisms \cite{varghese09}.
I hope that this work will help spur further effort along these lines.

The results presented in Fig.\ \ref{fig:btc} are a new analysis of earlier simulations \cite{hoy08} done in collaboration with Mark O. Robbins.
Stimulating discussions with Kenneth S.\ Schweizer, Leon Govaert, Mya Warren, and Mark Robbins, and support from NSF Award Nos.\ DMR-1006537 and DMR-0454947
are gratefully acknowledged.

\section{Appendix: Simulation Protocol}

All beads have mass $m$ and interact via the truncated and shifted Lennard-Jones potential
$U_{LJ}(r) = 4u_{0}[(a/r)^{12} - (a/r)^{6} - (a/r_{c})^{12} + (a/r_c)^{6}]$,
where $r_{c}=1.5a$ is the cutoff radius and $U_{LJ}(r) = 0$ for $r > r_{c}$.
The unit of time is $\tau_{LJ} = \sqrt{ma^{2}/u_{0}}$. 
Each polymer chain contains $N$ beads.
Covalent bonds are modeled using the finitely extensible nonlinear elastic (FENE) potential
$U(r) = -(1/2)(kR_{0}^2) {\rm ln}(1 - (r/R_{0})^{2})$, with $R_{0} = 1.5a$ and $k = 30u_{0}/a^{2}$\ \cite{kremer90}, and have variable length $\ell$, with the equilibrium value $\ell_0 = 0.96a$.
A bending potential $U_{bend}(\theta) = k_{bend}(1 - cos\theta)$,
where $\theta$ is the angle between consecutive covalent bond vectors along a
chain, imparts variable chain stiffness.
Two values of $k_{bend}$ are employed: flexible chains with $k_{bend} = 0$ have an entanglement length $N_e \simeq 85$, and semiflexible chains with $k_{bend} = 2u_0$ have $N_e \simeq 22$ \cite{everaers04}.  
The systems are well entangled; $N=500$ for flexible and $350$ for semiflexible chains.
Periodic boundaries are applied in all three directions, with cell dimensions $L_x$.  $L_y$, and  $L_z$ along the $x$, $y$, and $z$ directions.
Well-equilibrated melts are rapidly quenched to $k_B T = 0.01u_0 \sim T_g/35$. 
Uniaxial compression is applied along the $z$-direction at constant strain rate $\dot\epsilon = \dot{L}_z/L_z$, and $L_x$ and $L_y$ are varied to maintain zero stress along the transverse directions.

Simulation results were obtained using LAMMPS \cite{plimpton95}.
Bond damage (Fig.\ \ref{fig:btc}(b)) corresponds to local plastic rearrangements, identified with changes in intermonomer neighbor distances greater than 20\% over a strain interval $|\delta \epsilon|=.025$) \cite{hoy08}.  Nonaffine jumps (Fig.\ \ref{fig:btc}(d) and Eq.\ \ref{eq:d2dna}) are defined using the same $|\delta \epsilon|$.  The energy dissipated per damaged bond is given in units of $u_0$, and bond tensions (Fig.\ \ref{fig:btc}(c)) have units $u_0/a$.


\begin{thebibliography}{30}
\bibitem[{foo({\natexlab{a}})}]{footm}
\bibinfo{note}{An interesting recent result is the importance of covalent bond
  scission in systems deformed under extreme (shock) condition
  \cite{mattson10}. In these systems, scission will suppress long-range
  correlation of bond tensions and alter the cooperativity discussed below.}
\end{thebibliography}
\end{document}